\documentclass{elsart}

 \usepackage{graphicx}

\usepackage{amssymb}

\begin{document}

\begin{frontmatter}



\title{Nonvanishing string tension of elastic membrane models}

\author{Hiroshi Koibuchi},
\ead{koibuchi@mech.ibaraki-ct.ac.jp}
\author{Nobuyuki Kusano},
\author{Atsusi Nidaira},
\author{Komei Suzuki}

\address{Department of Mechanical and Systems Engineering, Ibaraki College of Technology, 
Nakane 866, Hitachinaka,  Ibaraki 312-8508, Japan}

\begin{abstract}
By using the grand canonical Monte Carlo simulations on spherical surfaces with two fixed vertices separated by the distance $L$, we find that the second-order phase transition changes to the first-order one when $L$ is sufficiently large. We find that string tension $\sigma \not= 0$ in the smooth phase while $\sigma \to 0$ in the wrinkled phase. 
 \end{abstract}

\begin{keyword}
Phase Transition \sep String Tension \sep Elastic Membranes 
\PACS  64.60.-i \sep 68.60.-p \sep 87.16.Dg
\end{keyword}
\end{frontmatter}

\section{Introduction}\label{intro}
In the past two decades, a considerable progress has been made \cite{David-book,Wiese-AP,Bowick-PR,WHEATER-rev,Sornette-Ostrowsky} on understanding the phase structure of Helfrich \cite{HELFRICH} and Polyakov-Kleinert \cite{POLYAKOV,Kleinert} model, which is an elastic surface model with bending rigidity denoted by $b$. The surface fluctuation can be viewed as a second order phase transition between the smooth phase at $b\!\to\!\infty$ and the crumpled phase at $b\!\to\!0$ \cite{Peliti-Leibler,DavidGuitter,David,BKS,BK,Kleinert-2}. Numerical studies have also been concentrated on the phase transition \cite{KANTOR-NELSON,KANTOR,GOMPPER-KROLL,WHEATER-cry,BCFTA,KY-2,BCHHM,ABGFHHM,CATTERALL,KOIB-PLA1,KOIB-PLA-2003-2,KOIB-PRE-2004-1}.

 The scaling of the string tension $\sigma$ was first investigated by Ambjorn et. al \cite{AMBJORN-Dur-Jon,AMBJORN-flu}.  It has been recognized that $\sigma$ vanishes at the critical point of the phase transition \cite{WHEATER-rev}.  The canonical Monte Carlo was used to extract information about $\sigma$. However, $\sigma$ is originally defined in the grand canonical ensemble. Therefore, it is interesting to use grand canonical Monte Carlo simulations in order to see the scaling of $\sigma$ more convincingly, although it has been demonstrated that the canonical MC can simulate the scaling of $\sigma$\cite{AMBJORN-flu}. 

It was recently reported that the phase structure of the model depends on the choice of the integration measure $\prod_i q_i^{\alpha} dX_i$, where $q_i$ is the co-ordination number of the vertex $i$ \cite{KOIB-PLA-2003-2}. From a conformal field theoretical viewpoint, this $\alpha$ is believed to be $2\alpha\!=\!3$ \cite{David-NP,BKKM,ADF,FN}. On the other hand, $q_i^{\alpha}$ is considered as a volume weight of the vertex $i$ in the integration $dX_i$. Hence, it is possible to extend $\alpha$ to continuous numbers by assuming that the weight can be chosen arbitrarily. Therefore, it is interesting to see the dependence of $\sigma$ on the phase transitions which can be controlled by the parameter $\alpha$. 

In this article, we would like to investigate the scaling property of $\sigma$.  It will be confirmed by using the grand canonical MC that the string tension vanishes at the continuous transition. Moreover, we would like to show a numerical evidence of the nonvanishing string tension in the smooth phase, which is distinguished from the wrinkled phase by the discontinuous transition at finite negative value of $\alpha$.    

We speculatively comment on why the result of nonvanishing string tension could be a relevant one. It is possible to consider that the nonvanishing string tension is connected to two interesting problems. The first is the problem of quark confinement, which is rather a problem in mathematical physics. The linear potential $V(L)\sim L$ assumed between quark and anti-quark separated by the distance $L$ gives the finite string tension, which is compatible with our result of nonvanishing string tension. The second is the conversion of external forces into an internal energy and vice versa in real physical membranes, and is a rather practical problem. If the model in this Letter represents properties in some real membranes, our result implies a possibility of such conversion.  

\section{The model}\label{model}
A sphere in ${\bf R}^3$ is discretized with piecewise linear triangles. Every vertex is connected to its neighboring vertices by bonds, which are the edges of triangles. Two vertices are fixed as the boundary points separated by the distance $L$.

The Gaussian energy $S_1$ and the bending energy $S_2$ are defined by
\begin{equation}
\label{S13-DISC}
S_1=\sum_{(ij)} \left(X_i-X_j\right)^2, \; S_2= \sum_i \left(1-\cos \theta_i\right), 
\end{equation}
where $\sum_{(ij)}$ is the sum over all bonds $(ij)$, and  $\theta_i$ in $S_2$ is the angle between two triangles sharing the edge $i$.

The partition function is defined by 
\begin{eqnarray}
 \label{Z-FLUID}
Z(b,\mu,\alpha; L) = \sum_N \sum_{T} \int \prod _{i=1}^N dX_i \exp\left[-S(X,{T})\right],\qquad \\
S(X,{T},N)=S_1 + b S_2 -\mu N - \alpha \sum_i \log q_i, \qquad\quad \nonumber
\end{eqnarray}
where $\sum_{T}$ denotes the sum over all possible triangulations ${T}$, $N$ is the total number of vertex. It should be noted that the chemical potential term $-\mu N$ and the co-ordination dependent term  $-\alpha \sum_i \log q_i$ are included in the Hamiltonian. The expression $S(X,{T},N)$  shows that $S$ explicitly depends on the variables $X$, ${T}$ and $N$. The coefficient $b$ is the bending rigidity, and $\mu$ is the chemical potential. $Z$ depends on $b$, $\mu$, $\alpha$, and $L$. The surfaces are allowed to self-intersect and hence phantom.

We expect 
\begin{equation}
\label{tension}
Z(b,\mu, \alpha; L)\sim \exp(-\sigma L)
\end{equation}
 in the limit $L\to \infty$ \cite{AMBJORN-flu}. Then, by using the scale invariance of the partition function, we have  \cite{WHEATER-rev,AMBJORN-flu}
\begin{equation}
\sigma = {2 \langle S_1\rangle - 3 \langle N\rangle \over L},
\end{equation}
where $\langle S_1\rangle$  and $\langle N\rangle$ are the mean values of $S_1$ and $N$.

The specific heat, which is the fluctuation of $S_2$, is defined by
  \\ $C_{S_2}\!=\!(b^2/\langle  N\rangle ) (\partial^2 \log Z / \partial b^2)$, and is calculated by using 
\begin{equation}
\label{Spec-Heat-S3}
C_{S_2} = {b^2\over \langle N\rangle} \langle \; \left( S_2 - \langle S_2 \rangle\right)^2 \; \rangle.
\end{equation}

\section{Monte Carlo technique}\label{MC-Techniques}
$X$ is updated so that $X^\prime \!=\! X \!+\! \delta X$, where the small change $\delta X$ is made at random in a small sphere centered at $X$. The radius $\delta r$ of the small sphere is chosen to maintain the rate of acceptance $r_X$ for the $X$-update as $0.5 \leq r_X \leq 0.55$. $\delta r$  is defined by using a constant number $\epsilon$ as an input parameter so that $\delta r \!=\! \epsilon\, \langle l\rangle$, where $\langle l\rangle$ is the mean value of bond length computed at every 250 MCS (Monte Carlo sweeps). It should be noted that $\delta r$ is almost fixed because $\langle l\rangle$ is constant and unchanged in the equilibrium configurations. 

${T}$ is updated by flipping a bond shared by two triangles. The bonds are labeled by sequential numbers and chosen randomly to be flipped. The rate of acceptance $r_{\it T}$ for the bond flip is uncontrollable, and the value of $r_{\it T}$ is about $30\% \leq r_{\it T} \leq 40 \%$. $N$-trials for the updates of $X$ and $N$-trials for ${T}$ are done consecutively and these make one MCS. 

$N$ is updated by both adsorption and desorption. In the desorption, a vertex is chosen at random, and then a bond that is connected to the vertex is chosen at random so that the two vertices at the ends of the bond unite and become a new vertex. In the adsorption, a triangle is chosen at random just like a bond chosen in the desorption, and a new vertex is added to the center of the triangle. As a consequence, the Euler number (=2) of the surface remains unchanged by the adsorption/desorption. The acceptance rate $r_N$ is uncontrollable as well as $r_{T}$, and the value of $r_N$ is about $55\% \leq r_N \leq 65\%$ in our MC. 

In the adsorption of a vertex, the corresponding change of the total energy ${\it \Delta}S\!=\!S({\rm new})\!-\!S({\rm old}) $ is calculated. The adsorption is then accepted with the probability ${\rm Min}[1,\exp\left(-{\it \Delta}S\right)/(N+1) ]$. In the desorption, ${\it \Delta}S\!=\!S({\rm new})\!-\!S({\rm old}) $ is calculated by assuming that one vertex is removed. The desorption is then accepted with the probability ${\rm Min}\left[1,N \exp\left(-{\it \Delta}S\right)\right]$. The adsorption/desorption are tried alternately at every 5 MCS.

 We use surfaces of size $N\!\simeq\!500$, $N\!\simeq\!1000$, and $N\!\simeq\!1500$.  $N$ depends on both $\mu$ and $\alpha$ which is fixed to $\alpha\!=\!-5.5$, and $N$ is almost independent of $L$. There is no a piori reason for choosing $\alpha\!=\!-5.5$, although we have confirmed that the phase transition occurs at $\alpha\!\simeq\!-5.5$ in the model with the fixed center of surface. The values of $\mu$ are chosen so that $N\!\simeq\!500$, $N\!\simeq\!1000$, and $N\!\simeq\!1500$. The diameter $L_0(N)$ of the spheres at the start is fixed so that $\sum_i l_i^2\!\simeq\!3N/2$, where $l_i$ is the length of the bond $i$. As a consequence, $L_0(N)$ becomes
\begin{equation}
\label{L-scale}
L_0(N) \propto  \sqrt{N}.
\end{equation}
 We use two kinds of $L$ for each $L_0(N)$ so that $L\!=\!1.5L_0(N)$ and $L\!=\!3L_0(N)$. The distance of the boundary points is increased from $L_0(N)$ to $L$ in the first $5\!\times\! 10^6$ MCS.

It should be noted that both $L\!=\!1.5L_0(N)$ and $L\!=\!3L_0(N)$ becomes $\infty$ in the thermodynamic limit $N\!\to\!\infty$ because of Eq.(\ref{L-scale}). Therefore, $\sigma$ defined by Eq.(\ref{tension}) can be extracted from these values of $L$ at sufficiently large $N$.

\section{Results}\label{Results}
\begin{figure}[hbt]
\centering
\includegraphics[width=10cm]{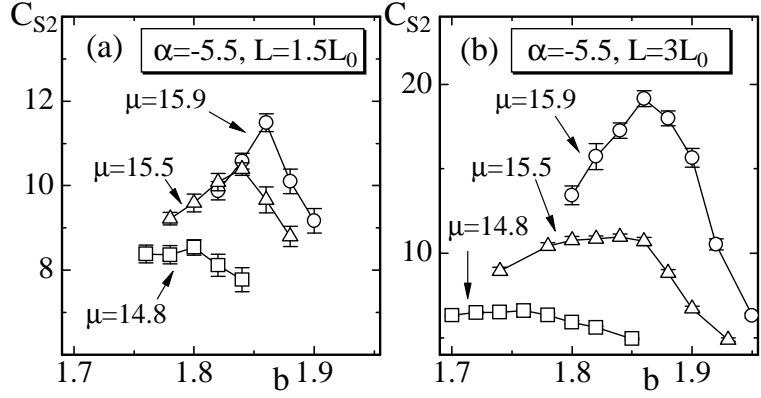}
\caption{$C_{S_2}$ vs $b$ at (a) $\alpha\!=\!-5.5$, $L\!=\!1.5L_0(N)$  and (b) $\alpha\!=\!-5.5$, $L\!=\!3L_0(N)$. $\mu\!=\!15.9$ ($\bigcirc$), $\mu\!=\!15.5$ ($\triangle$), and $\mu\!=\!14.8$ ($\Box$) respectively correspond to $N\!\simeq\!1500$, $N\!\simeq\!1000$, and $N\!\simeq\!500$.} 
\label{fig-1}
\end{figure}
The specific heat $C_{S_2}$ at $\alpha\!=\!-5.5$ is plotted against $b$ in Figs. \ref{fig-1}(a) and \ref{fig-1}(b). The number of vertices is $N\!\simeq\!500(\Box)$, $N\!\simeq\!1000(\triangle)$, and $N\!\simeq\!1500(\bigcirc)$ in each figure. We find that the peak value $C_{S_2}^{\rm max} $ grows with increasing $N$. An interesting point to emphasize is that the phase transition is strengthened when $L$ is increased from $L\!=\!1.5L_0(N)$ to $L\!=\!3L_0(N)$. 

\begin{figure}[hbt]
\centering
\includegraphics[width=10cm]{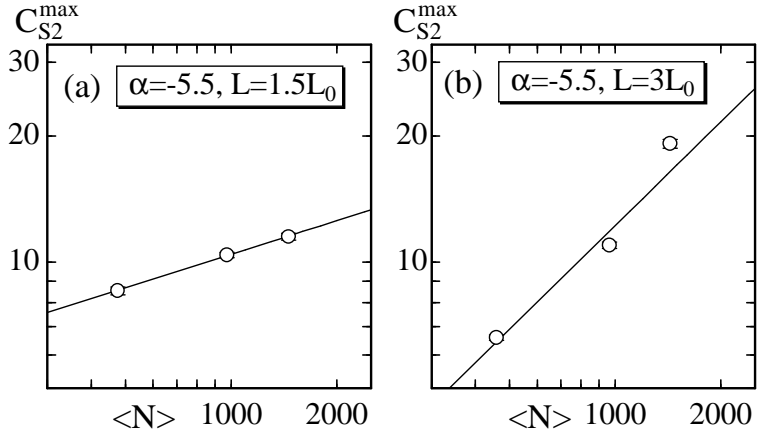}
\caption{$C_{S_2}^{\rm max}$ vs $\langle N\rangle$ in log-log scale at (a) $\alpha\!=\!-5.5, L\!=\!1.5L_0$, and at (b) $\alpha\!=\!-5.5, L\!=\!3L_0$. }
\label{fig-2}
\end{figure}
To see the order of the transition, we plot $C_{S_2}^{\rm max}$ vs $\langle N\rangle$ in Figs. \ref{fig-2}(a) and \ref{fig-2}(b) in log-log scale. We clearly find in both of the figures that $C_{S_2}^{\rm max}$ scales according to $C_{S_2}^{\rm max}\!\sim\!N^\nu$, and we have 
\begin{eqnarray}
\label{nu}
&&\nu=0.265\pm 0.025 \quad \left[\alpha=-5.5, \; L=1.5L_0(N)\right], \nonumber \\
&&\nu=0.822\pm0.182 \quad \left[\alpha=-5.5,\; L=3L_0(N)\right]. 
\end{eqnarray}
 The value $\nu\!=\!0.265(25)$ suggests that the phase transition is of second order at $\alpha\!=\!-5.5$ when $L\!=\!1.5L_0(N)$.  On the other hand at $\alpha\!=\!-5.5$, $L\!=\!3L_0(N)$, the value $\nu\!=\!0.822(182)$ is still slightly smaller than $\nu\!=\!1$. Nevertheless, the result implies that the model undergoes the discontinuous transition at $\alpha\!=\!-5.5$ when $L\!=\!3L_0(N)$. 

\begin{figure}[hbt]
\centering
\includegraphics[width=10cm]{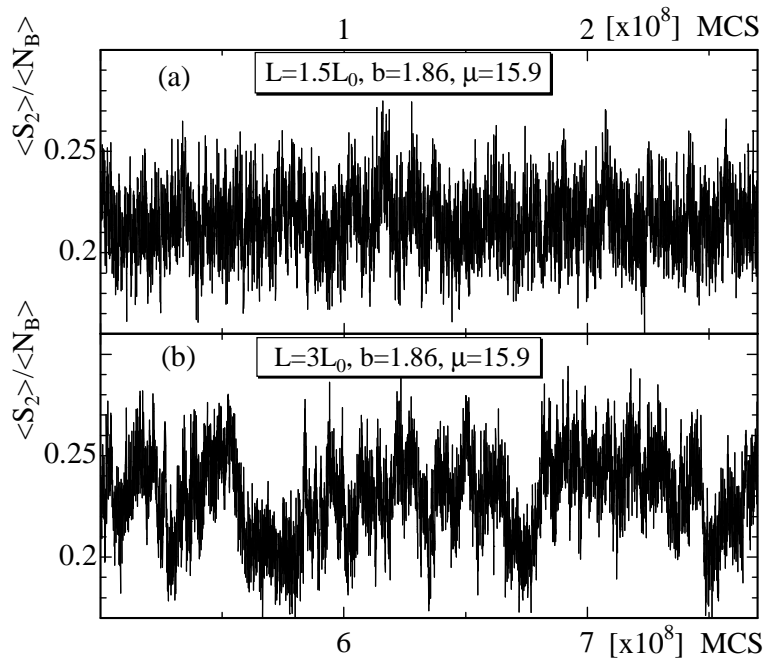}
\caption{Variation of $\langle S_2\rangle /\langle N_B\rangle $ against MCS at (a) $b\!=\!1.86, L\!=\!1.5L_0$ (at the critical point of the continuous transition) and at (b) $b\!=\!1.86, L\!=\!3L_0$ (at the discontinuous transition point), where $\alpha\!=\!-5.5$, $\mu\!=\!15.9$.}
\label{fig-3}
\end{figure}
To see the difference between the transitions shown in Figs. \ref{fig-2}(a) and \ref{fig-2}(b), we depict $\langle S_2\rangle/\langle N_B\rangle$ against MCS in Figs. \ref{fig-3}(a) and \ref{fig-3}(b). These were obtained at $b\!=\!1.86, L\!=\!1.5L_0, \mu\!=\!15.9$ and at $b\!=\!1.86, L\!=\!3L_0, \mu\!=\!15.9$, where $N_B$ is the total number of bond. It is possible to see in Fig. \ref{fig-3}(b) that there are two distinct states which differ in values of $S_2$; one is characterized by $\langle S_2\rangle/\langle N_B\rangle\!\simeq\! 0.2$ and the other by  $\langle S_2\rangle/\langle N_B\rangle\!\simeq\! 0.25$.

Large number of MCS was needed to obtain $\nu$ in Eq. (\ref{nu}). $9.5\!\times\!10^8$ MCS were done for $N\!\simeq\!1500$ surfaces at the vicinity of the transition point when $L\!=\!3L_0(N)$. Relatively small number of MCS was done when $L\!=\!1.5L_0(N)$. 

\begin{figure}[hbt]
\centering
\includegraphics[width=10cm]{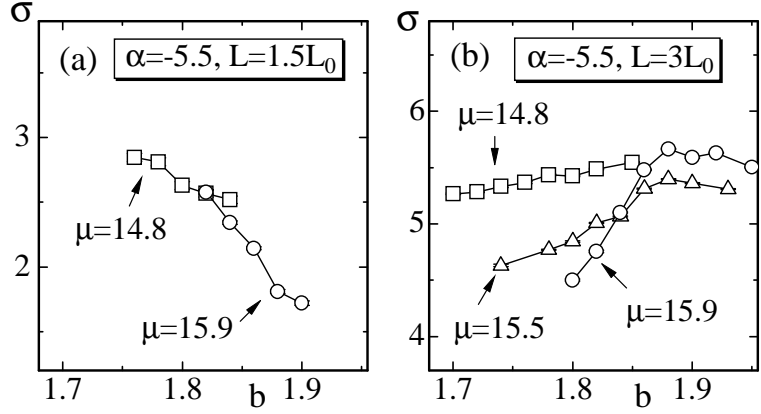}
\caption{$\sigma$ vs $b$ obtained at (a) $\alpha\!=\!-5.5$, $L\!=\!1.5L_0$ and at (b) $\alpha\!=\!-5.5$, $L\!=\!3L_0$.}
\label{fig-4}
\end{figure}
The dependence of $\sigma$ on $b$ is plotted in Figs. \ref{fig-4}(a), \ref{fig-4}(b), where the symbols ($\bigcirc, \triangle, \Box$) in the figures correspond to those in Figs. \ref{fig-1}(a), \ref{fig-1}(b). We immediately understand from the figures that the dependence of $\sigma$ on $b$ in Fig. \ref{fig-4}(a) is completely different from that in Fig. \ref{fig-4}(b). $\sigma$ decreases with increasing $b$ when $L\!=\!1.5L_0(N)$. On the contrary, $\sigma$ increases with increasing $b$ when $L\!=\!3L_0(N)$. Moreover, we find in Fig. \ref{fig-4}(b) that $\sigma\!\to\!\sigma_0({\rm const})$ in the smooth phase close at $b_c(\mu,\alpha)$, and that $\sigma_0$ is independent of $N$, where $b_c(\mu,\alpha)$ denotes the value of $b$ where $C_{S_2}$ has the peak as shown in Fig. \ref{fig-1}(b), and denotes that $b_c$ depends on $\mu$ and $\alpha$.  Figure \ref{fig-4}(b) also shows that $\sigma $ rapidly changes against $b$ at $b_c(\mu,\alpha)$ when $N$ (or $\mu$) increases.

In order to show the dependence of $\sigma$ on $\langle N\rangle$ more clearly, we introduce the reduced bending rigidity 
\begin{equation}
\lambda = {b \over b_c(\mu,\alpha)} - 1.
\end{equation}
Then, the transition point $b_c(\mu,\alpha)$ is represented by $\lambda\!=\!0$, the smooth phase at $b\!>\!b_c(\mu,\alpha)$ by $\lambda\!>\!0$, and the wrinkled phase at $b\!<\!b_c(\mu,\alpha)$ by $\lambda\!<\!0$. 

\begin{figure}[hbt]
\centering
\includegraphics[width=10cm]{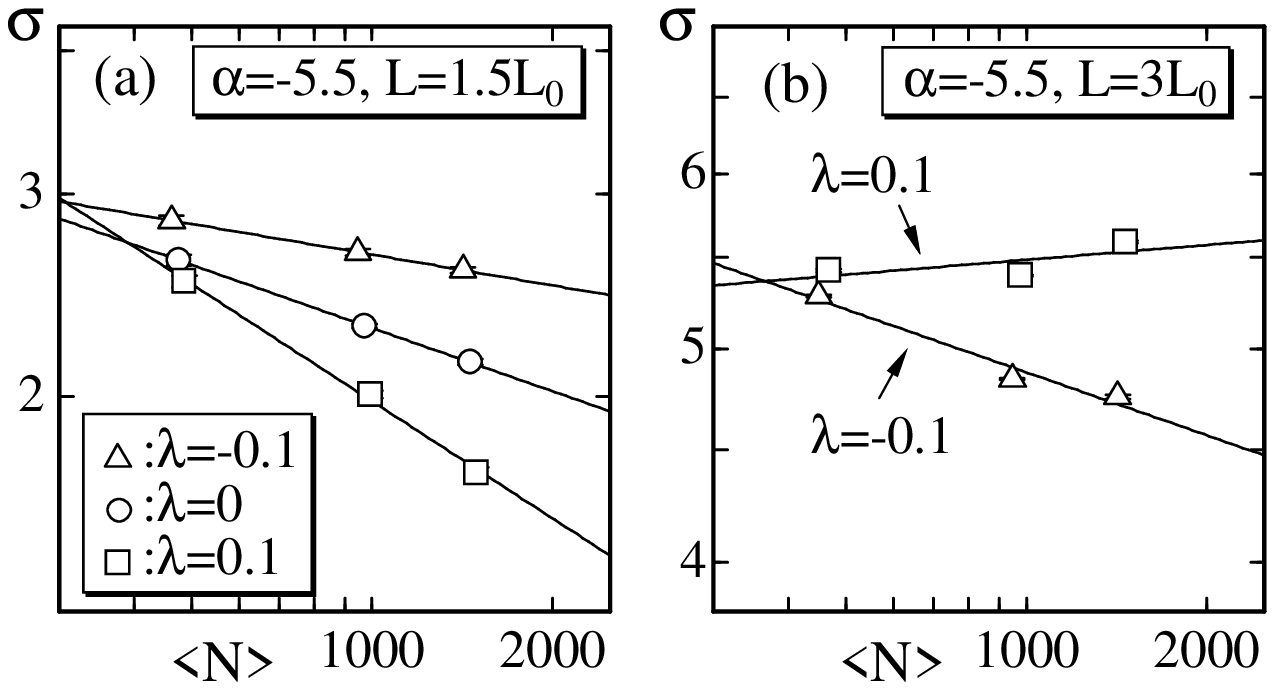}
\caption{$\sigma$ vs $\langle N\rangle$ in log-log scale obtained at (a) $\alpha\!=\!-5.5$, $L\!=\!1.5L_0$ and at (b) $\alpha\!=\!-5.5$, $L\!=\!3L_0$. }
\label{fig-5}
\end{figure}
Figures \ref{fig-5}(a), \ref{fig-5}(b) show log-log plots of $\sigma$ against $\langle N\rangle$ obtained at $\lambda\!=\!-0.1(\triangle)$, $\lambda\!=\!0(\bigcirc)$, and $\lambda\!=\!0.1(\Box)$. The straight lines in each figure denote the scaling property of $\sigma$ such as
\begin{equation}
\label{sigma-scale}
\sigma \propto N^{-\kappa} \qquad(\kappa \geq 0).
\end{equation}
It should be emphasized that the scaling in Eq. (\ref{sigma-scale}) is compatible with  $\sigma \!\propto\! (L/N)^{\delta} $  in \cite{AMBJORN-flu},  since $L\!\propto\! L_0(N)\!\propto\! \sqrt{N}$ as described in Eq. (\ref{L-scale}). We  find in Figs. \ref{fig-5}(a) that $\sigma\!\to\!0(N\!\to\!\infty)$, which is the scaling property at the continuous transition in \cite{AMBJORN-flu}.

On the contrary, we clearly see in Fig. \ref{fig-5}(b) that
\begin{equation}
\label{sigma-0}
 \sigma\to\sigma_0({\rm const}), \quad \left[b> b_c(\mu,\alpha)\right].
\end{equation}
This corresponds $\kappa\!\to\!0$ in Eq. (\ref{sigma-scale}), and is the main result of this Letter. Moreover, we see $\sigma\!\to\!0(N\!\to\!\infty)$ in the wrinkled phase at $b\!<\! b_c(\mu,\alpha)$ just like $\sigma$ at the continuous transition seen in Fig. \ref{fig-5}(a). It should be noted that the scaling of $\sigma$ was not seen at the discontinuous transition point, where physical quantities might be ill-defined.

\begin{figure}[hbt]
\centering
\includegraphics[width=10cm]{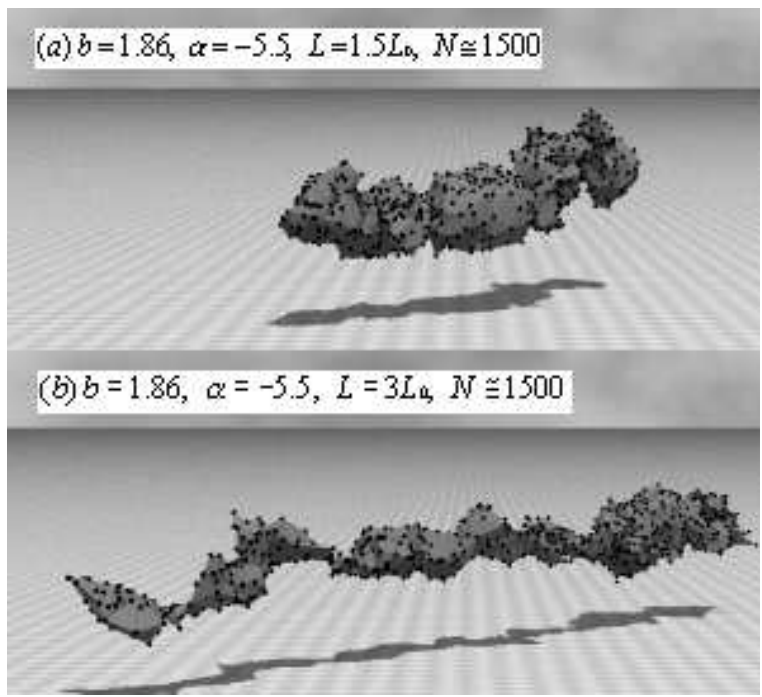}
\caption{Snapshots of surfaces at (a) $b\!=\!1.86$, $\alpha\!=\!-5.5$, $L\!=\!1.5L_0$ and at (b) $b\!=\!1.86$, $\alpha\!=\!-5.5$, $L\!=\!3L_0$. }
\label{fig-6}
\end{figure}
Snapshots of surfaces are shown in Figs. \ref{fig-6}(a) and \ref{fig-6}(b), which are obtained at $b\!=\!1.86$, $\alpha\!=\!-5.5$. The size is $N\!\simeq\!1500$ in each surface. The surface in Fig. \ref{fig-6}(a) looks short and swollen, whereas the surface in Fig. \ref{fig-6}(b) long and slight, as expected. 

\section{Summary and conclusion}\label{Conclusions}
To summarize the results, we have studied the scaling properties of string tension of elastic membranes by grand canonical MC. The model contains a measure term $-\alpha \sum_i \log \sigma_i$ in the Hamiltonian. The parameter $\alpha$ was assumed as a continuous one and fixed to $\alpha\!=\!-5.5$. The chemical potential $\mu$ were chosen so that $N\!\simeq\! 500$, $N\!\simeq\! 1000$, and $N\!\simeq\! 1500$. Spherical surfaces were stretched, and then two boundary vertices were separated by length $L\!=\!1.5L_0(N), 3L_0(N)$, where $L_0(N)(\propto\sqrt{N})$ is the diameter of the sphere for the starting configuration of MC. The model undergoes the first-order phase transition on the surfaces of $L\!=\!3L_0(N)$ at $\alpha\!=\!-5.5$. It was found that $\sigma$ becomes finite in the smooth phase at $b\!> \!b_c(\mu,\alpha)$ and vanishes in the wrinkled phase at $b\!< \!b_c(\mu,\alpha)$. When the length $L$ is reduced to $L\!=\!1.5L_0(N)$, the phase transition changes to the second-order one, and $\sigma$ vanishes at the vicinity of the critical point of the continuous transition in the limit $N\!\to\! \infty$ as expected. 

We have explored the scaling of $\sigma$ at $\alpha\!=\!-5.5, 0.0, 5.5$ with $L\!=\!1.5L_0(N)$, $2L_0(N), 3L_0(N)$, including the case presented in this Letter. Non-vanishing $\sigma$ was found only in the case $\alpha\!=\!-5.5$ with $L\!=\!3L_0(N)$. More detailed information on the simulation data will be presented elsewhere. 

This work is supported in part by the Grant-in-Aid for Scientific Research 15560160.



\end{document}